\documentclass[hyper,12pt,letterpaper]{JHEP3}
\usepackage{epsfig,graphicx,bm,amsmath}
\usepackage{showlabels}
\newcommand{\be}{\begin{equation}}
\newcommand{\ee}{\end{equation}}
\newcommand{\bea}{\begin{eqnarray}}
\newcommand{\eea}{\end{eqnarray}}
\newcommand{\beq}{\begin{equation}}
\newcommand{\eeq}{\end{equation}}
\newcommand{\beqa}{\begin{eqnarray}}
\newcommand{\eeqa}{\end{eqnarray}}
\newcommand{\beqar}{\begin{eqnarray*}}
\newcommand{\eeqar}{\end{eqnarray*}}

\newcommand{\beas}{\begin{eqnarray*}}
\newcommand{\eeas}{\end{eqnarray*}}
\usepackage{epsf}

\title{
{\bf Higher derivative correction to Kaluza-Klein black hole solution}
}
\bigskip
\author{
 Hossein Yavartanoo $^1$\thanks{yavar [at] phya.snu.ac.kr}
 ~and
 Sangheon Yun $^2$\thanks{sanhan1 [at] phya.snu.ac.kr}\\
~\\
$^1${\it\small Center for Theoretical Physics and BK-21 Frontier Physics Division,
 Seoul National University, Seoul 151-747 KOREA} \\
\\
$^2${\it\small School of Physics and Astronomy, Seoul National University,
 Seoul 151-747 KOREA  }\\}

\date{}

\maketitle
\newpage
\abstract{We investigate the attractor mechanism in Kaluza-Klein black hole solution
in the presence of higher derivative terms. In particular, we discuss
the attractor behavior of static black holes by using the effective potential approach
as well as entropy function formalism. We consider different higher derivative terms with a general coupling
to moduli field.
For the $R^2$ theory, we use effective potential approach, looking for solutions which are
analytic near the horizon and show that they exist and enjoy the attractor behavior.
The attractor point is determined by extremization of the modified effective potential
at the horizon. We study the effect of the general higher derivative corrections of $R^n$ terms.
Using the entropy function we define the modified effective potential and we find the conditions
to have the attractor solution. In particular for a single charged Kaluza-Klein black hole solution we show that higher derivative correction dresses the naked singularity for an appropriate coupling, and we can find the attractor solution.

}

\preprint{}


\begin{document}
\vfill \setcounter{page}{0} \setcounter{footnote}{0}

\section{Introduction}

One of the main achievements of string theory has been a successful explanation of black hole entropy \cite{Strominger:1996sh}. For a large class of BPS black hole solutions, we are able to identify the black hole entropy as the logarithm of the degeneracy of states belonging to the microstates of the corresponding black hole. This prediction includes not only the leading entropy formula of Bekenstein and Hawking (see \cite{Mohaupt:2007mb} for a recent review) but also all the subleading quantum gravitational corrections which was proposed in \cite{Ooguri:2004zv}, building on the work of \cite{Lopes Cardoso:1998wt} (see also \cite{Mohaupt:2005jd} and references there).

\bigskip

BPS black holes are known to exhibit an attractor mechanism, whereby the values of scalar fields at the horizon are determined only in terms of the charges carried by the black hole and are independent of the asymptotic values of the scalar fields. As a result, the black hole solution at the horizon and the resulting entropy turn out to be determined completely in terms of the conserved charges. This phenomenon plays an important role in understanding the entropy of asymptotically flat non-supersymmetric extremal black holes in string theory \cite{Dabholkar:2006tb}.

\bigskip

This phenomenon has been discovered in the context of ${\mathcal N}=2$ supergravity \cite{Ferrara:1995ih}, then extended to other supergravity theories \cite{Ferrara:1996um}. It is well understood that supersymmetry does not really play an essential role in the attractor mechanism. The attractor phenomenon works as a consequence of the enhanced
symmetry of the near horizon geometry of an extremal black hole, which is $AdS_2\times S^p$ for a static spherically symmetric black hole in $p+2$ dimensions \cite{Sen:2005wa}. In fact, the `long throat' of $AdS_2$ is at the basis of attractor mechanism \cite{Sen:2005wa, Astefanesei:2006dd}\footnote{A relation between the
entanglement entropy of dual conformal quantum mechanics in $AdS_2/CFT_1$ and the
entropy of an extremal black hole was provided in \cite{Azeyanagi:2007bj}.}.

\bigskip

Over the last few decades, Kaluza-Klein black holes have attracted considerable attention. The original Kaluza-Klein black hole solution in four dimensions was obtained by compactification of five-dimensional pure gravity on a circle \cite{Rasheed:1995zv}. The field contents of this theory are a $U(1)$ gauge field, a scalar field, and gravity. Stationary black hole solutions are parametrized by their electric $Q$ and magnetic $P$ gauge
charges, as well as their mass $M$ and angular momentum $J$. There is a simple embedding of the system into string theory. Taking the product of the original five dimensional solution with a flat $T^6$ (or a general Calabi-Yau three-fold), we obtained a solution to $M$ theory, whose $IIA$ reduction has $D0$ and $D6$ charges. Even in the extremal limit, this black hole is not supersymmetric \cite{Khuri:1995xq}, due to the absence of supersymmetric bound states of $D0$ and $D6$ branes. However, there is a quadratically stable non-supersymmetric, $D0-D6$ bound states \cite{Taylor:1997}, and this will serve as a basis to the microscopic picture, which have been discussed in \cite{Emparan:2006it}. Having a successful statistical description of the leading entropy formula of the Kaluza-Klein black hole, one natural question is whether one can consider the subleading quantum gravitational corrections in this picture. 

\bigskip

During the last several years, study of higher derivative corrections to the entropy of
supersymmetric black holes has provided successful results in string theory. In many examples these corrections match the appropriate corrections to the statistical entropy of the corresponding microscopic system. 
While studying the higher derivative corrections to the entropy of a generic black
hole is a difficult problem, a general method for computing the entropy of extremal black holes was developed in \cite{Sen:2005wa}. This method does not provide an explicit construction of the full black hole solution, but it gives a way to compute the near horizon field configuration and entropy of an extremal, but not necessarily supersymmetric black hole, with a given set of charges.

\bigskip

In this paper we examine the effect of higher derivative corrections to the Kaluza-Klein black hole near horizon solution. Following \cite{Goldstein:2005hq} and \cite{yavar} (for some related works, see also \cite{Alishahiha:2006ke}), we use perturbative methods and numerical analysis to show that, the horizon of extremal Kaluza-Klein black hole in the presence of a general higher derivative correction $R^n$, is attractor. This analysis supports the existence of the attractor mechanism for Kaluza-Klein black hole with higher derivative corrections.

\bigskip

We start with a simple example of such corrections, Gauss-Bonnet correction. Although in four dimensions this term is a total derivative, when it coupled to moduli field, it can not be integrated out. To study the effect of this term in attractor solution, we use the near horizon analysis \cite{Goldstein:2005hq}. We will show that the condition to have attractor solution are succinctly stated in terms of a ''modified effective potential'' $W$ for the scalar field. The condition is that, $W(\phi)$ as a function of the moduli field, must have a minimum, which means $\partial  W(\phi_0) =0$, and $\partial^2 W(\phi_0)$ is positive. The resulting attractor value for the moduli is the critical value, $\phi_0$. Moreover, the entropy of the black hole is proportional to $W(\phi_0)$, and is thus independent of the asymptotic values for the moduli. By using the entropy function method we will find the modified effective potential for the general $R^2$ correction. We also study higher derivative corrections of the form $R^n$ as well and we will introduce the modified effective potential for these general higher derivative corrections.

\bigskip

Although in this paper we are mainly interested in Kaluza-Klein black holes, but the results we have found is applicable to all extremal black hole solutions in Einstein-Hilbert gravity coupled to arbitrary number of moduli fields and $U(1)$ gauge fields in the presence of higher derivative corrections. The only difference is we need to use general form for the effective potential $V_{eff}$, introduced in \cite{Goldstein:2005hq}.

\bigskip

The paper is organized as follows: In section 2 we review Kaluza-Klein Black hole solution. In section 3 we will consider $R^2$ correction to Kaluza-Klein black hole solution. We restrict ourselves to Gauss-Bonnet term. Using entropy function method, we generalize our results to a general $R^2$ correction. In section 4 we study general $R^n$ corrections. The last section is devoted to discussion.

\section{Review of Kaluza-Klein black hole solution}

We consider four-dimensional Kaluza-Klein black hole solution. This solution has been found by reducing five-dimensional Kerr black hole to three dimensions, applying $SO(2,1)$ boosts and oxidizing up to four dimensions. It is the solution of five-dimensional Einstein-Hilbert action reduced to four dimensions:
\be
\label{action0}
I = \frac{1}{16 \pi G_4} \int d^4x \sqrt{-G_{(4)}}\left( R_{(4)}-\frac{2}{3}\partial^{\mu}\Phi \partial_{\mu}\Phi- e^{2\Phi}F^{\mu\nu}F_{\mu\nu}\right)
\ee
The original five-dimensional metric is given by
\be
ds_{(5)}^2= e^{4\Phi/3} (dy +4 A_{\mu} dx^{\mu})^2 + e^{-2 \Phi/3} ds_{(4)}^2
\ee
and the four-dimensional metric is
\be
ds^{2}_{(4)} = -\frac{H_3}{\sqrt{H_1H_2}} (dt+ \omega d\phi)^2 +\frac{\sqrt{H_1H_2}}{\Delta} d\rho^2 + \sqrt{H_1H_2} d\theta^2 + \frac{\Delta \sqrt{H_1H_2}}{H_3} \sin^2\theta d\phi^2
\ee
where
\bea
H_{1} &=& \rho^{2}+a^{2}\cos^{2}\theta+\rho(p-2m)+{p\over
p+q}{(p-2m)(q-2m)\over 2} \nonumber \\ &~&\qquad - {p\over 2m(p+q)}
\sqrt{(p^{2}-4m^{2})(q^{2}-4m^{2})}~a\cos\theta~,\\
H_{2} &=& \rho^{2}+a^{2}\cos^{2}\theta+\rho(q-2m)+{q\over
p+q}{(p-2m)(q-2m)\over 2} \nonumber \\ &~&\qquad +
{q\over 2m(p+q)}
\sqrt{(p^{2}-4m^{2})(q^{2}-4m^{2})}~a\cos\theta~,\\
H_{3} &=& \rho^{2}+a^{2}\cos^{2}\theta-2m\rho~,\\
\Delta &=& \rho^{2}+a^{2}-2m\rho~, \\
\omega &=& \sqrt{pq}{(pq+4m^{2})\rho-m(p-2m)(q-2m)\over 2m(p+q)H_{3}}~a
\sin^{2}\theta d\phi~,
\eea
The solution for the dilaton is given by
\be
\label{dilaton}
e^{4\Phi/3} = \frac{H_2}{H_1}
\ee
where the dilaton has been asymptotically set to zero. Finally the gauge field is given by
\bea
A_{\mu} dx^{\mu}&=& -\left[ 2Q(\rho+{p-2m\over 2}) +
\sqrt{q^{3}(p^{2}-4m^{2})\over 4(p+q)}\frac{a\cos\theta}{m}\right] H_{2}^{-1} dt
\nonumber \\
&~&\qquad-H_{2}^{-1} \left[2P(H_{2}+ a^{2}\sin^{2}\theta)\cos\theta+
\sqrt{p(q^{2}-4m^{2})\over
4(p+q)^{3}}\times \right.
\nonumber \\ &~&
\left.\phantom{\sqrt{p q^{2})\over
4m^{2}}} \times
\left[(p+q)(p\rho-m(p-2m))+q(p^{2}-4m^{2})\right]\frac{a\sin^{2}\theta}{m}
\right]d\phi~, \;\;\;\;\;\;\;\;\;
\eea
The four parameters $(m,a,q,p)$ appearing in the solution are related to
the physical mass $M$, angular momentum $J$, electric charge $Q$, and magnetic
charge $P$ through :
\bea
2G_{4}M&=& {p+q\over 2}~,
\label{eq:paraM} \\
G_{4}J &=& {\sqrt{pq}(pq+4m^2)\over 4(p+q)}~a~,
\label{eq:paraJ} \\
Q^{2} &=& {q(q^2-4m^2)\over 4(p+q)}~,
\label{eq:paraQ} \\
P^{2} &=& {p(p^2-4 m^2)\over 4(p+q)}~.
\label{eq:paraP}
\eea

By eliminating conical singularity in the Euclidean $(\tau= it,~\rho)$ sector, we obtain the temperature
\be
\label{T}
T = \frac{m}{\pi \sqrt{pq}} \left[ \frac{pq +4m^2}{p+q} + \frac{2m^2}{\sqrt{m^2-a^2}} \right]^{-1},
\ee
where $P$ and $Q$ are electric and magnetic charges carried by black hole. The area of the black hole can be determined from the four-dimensional Einstein metric and it gives the black hole entropy:

\be
S = \frac{A}{4 G_4} = {\pi{pq}\over G_{4}}
\left[ m + {pq+4m^2\over 2m(p+q)}\sqrt{
m^2-a^2} \right]\;.
\ee

Next we note that the action (\ref{action0}) has a scaling symmetry:
\be
\Phi\rightarrow \Phi + \Phi_{\infty}\, , \;\;\;\;  F_{\mu\nu}\rightarrow   e^{\Phi_{\infty}} F_{\mu\nu},
\ee
for a constant $\Phi_{\infty}$. Therefore we can generate one parameter family of solutions carrying fixed electric and magnetic charges by using the transformation:
\be
\Phi \rightarrow \Phi+ \Phi_{\infty}, \;\;\; F\rightarrow e^{-\Phi_{\infty}} F, \;\;\; Q \rightarrow e^{\Phi_{\infty}} Q, \;\;\; P \rightarrow e^{-\Phi_{\infty}} P
\ee
The horizon and ergoregion are given by:
\bea
&&\Delta=0 \Longrightarrow \rho_{\pm}= m\pm \sqrt{m^2-a^2} \cr\crcr
&&\hspace{10mm} g_{tt}= -\frac{H_3}{\sqrt{H_1H_2}}=0 \nonumber
\eea
The extremal limit is defined as the limit of degenerate horizon and zero temperature. From (\ref{T}) it is clear that it can be achieved in two ways, going to two distinct branches of solution:
\begin{itemize}
 \item {\bf Ergo-free branch:}
\end{itemize}

Consider the limit: $m, a \rightarrow 0$ with $a/m, p$ and $q$ held finite. The horizon is located at $\rho=0$.
In this limit $p, q$ and $a/m$ can be taken as the independent parameters labeling the solution. In this limit it is easy to see that the angular velocity of the horizon vanishes and there is no ergosphere. The mass and the black hole entropy in this case are given by:
\be
2G_4M = (Q^{2/3}+P^{2/3})^{3/2}\, ,\;\;\;\;\; S= 2\pi \sqrt{\frac{P^2Q^2}{G_4^2}-J^2}
\ee
where $G_4|J| < |PQ|$
\begin{itemize}
 \item {\bf Ergo branch:}
\end{itemize}
The extremal limit on this branch is given by taking $a=m$ in the black hole solution. The horizon is located at $\rho=m$. In this case we note that $H_3$ changes from being positive at large distance to negative at horizon, therefore $g_{tt}$ changes sign as we go from the asymptotic region to the horizon and the solution has an ergosphere. In this branch the entropy is given by
\be
S= 2\pi \sqrt{J^2 - \frac{P^2Q^2}{G_4^2}}
\ee
where $G_4|J| > |PQ|$.

At the dividing value  $G_4|J| = |PQ|$ the extremal horizon disappears and the solution has a naked singularity.

\section{$R^2$ correction to Kaluza-Klein Black Hole solution}

In this section we use the results of \cite{Goldstein:2005hq} to study attractor mechanism for Kaluza-Klein black hole solution with a general $R^2$ correction. We also discuss
the attractor mechanism using the entropy function framework
\cite{Sen:2005wa}. The entropy function approach is based on the
near-horizon geometry and its enhanced symmetries. The analysis in \cite{Goldstein:2005hq} is based on investigating
the equations of motion of the moduli and finding the conditions
satisfied by the effective potential such that the attractor
phenomenon occurs. We also use the entropy function method to find a general $R^n$ correction in the next section.

\subsection{Effective potential and non-supersymmetric attractor}
\label{veff}

In this section, we consider four-dimensional, extremal, nonrotating Kaluza-Klein black hole solution with $R^2$ correction. Then the action is given by
\bea
\label{action1}
I = \frac{1}{16 \pi G_4} \int d^4x \sqrt{-G_{(4)}}\bigg( R_{(4)}-\frac{2}{3}\partial^{\mu}\Phi \partial_{\mu}\Phi- e^{2\Phi}F^{\mu\nu}F_{\mu\nu} +   {\mathcal L}_{2}\bigg)
\eea

where ${\mathcal L}_{2}$ is a general $R^2$ correction with the following form

\bea
\label{R2c}
 {\mathcal L}_2= G(\phi)\left(\alpha R_{}^2 + \beta R_{\mu\nu} R^{\mu\nu} + \gamma R_{\mu\nu\rho\sigma} R^{\mu\nu\rho\sigma}\right) \;.
\eea

$\alpha, \beta$ and $\gamma$ are arbitrary constants and $G(\Phi)$ is a dilaton dependent coupling. Our primary interest is in the study of the case when the ${\mathcal L}_2$ part of the
action (\ref{action1}) forms the Gauss-Bonnet combination
\be
\alpha=1\; ,\,\,\,\, \beta=-4 \; ,\,\,\,\, \gamma=1
\ee
Theory with Gauss-Bonnet combination is a very special case of higher derivative gravity. Since the Gauss-Bonnet term in four
dimensions is a topological term,
the coupling to the scalar fields is crucial. Without this coupling, the Gauss-Bonnet
term does not contribute to the equations of motion. In that case, we are left with
the KK solution reviewed in previous section.

In this section we use the result of \cite{Goldstein:2005hq,Alishahiha:2006ke} to study the Kaluza-Klein black hole solution and we discuss the attractor mechanism in the presence of  $R^2$ corrections. We investigate the equations of motion of the moduli and find the conditions satisfied by the modified effective potential such that the attractor phenomenon occurs.

We focus on static and spherically symmetric space-time metric which can be written as:
\be
ds^2=-a(r)^2dt^2+\frac{dr^2}{c(r)^2} + b(r)^2 \left(d\theta^2+\sin^2\theta d\varphi^2\right)
\ee

Although it is possible to set $c(r) = a(r)$ by redefinition of the radial coordinate, we
keep $c(r)$ free to derive the complete set of equations, including the Hamiltonian
constraint, from the one-dimensional action.

Using the above ansatz, one finds the following one-dimensional action:

\bea
\label{action2}
I = \frac{1}{2 G_4} \int dt dr \bigg[&-& \hspace{-1mm}b^2(a'c)' + \frac{a}{c}(1 - c^2 b'^2) - 2b(acb')' - \frac{1}{3} acb^2\phi'^2 - \frac{a}{cb^2}V_{eff}(\phi) \cr
&+& \hspace{-1mm}
4 G(\phi) \left(c a'(-1+ c^2 b'^2)\right)' \bigg]
\eea
where $V_{eff}(\phi)$ is given by
\be
V_{eff} (\phi) = e^{-2\phi}Q^2 + e^{2\phi}P^2
\ee

For the action (\ref{action1}) the Bianchi identity and equation of motion for the gauge field can be solved by a field strength of the form
\be
F=\frac{Q}{b^2} dt \wedge dr + P \sin\theta d\theta \wedge d\varphi \; ,
\ee
where $Q$ and $P$ are electric and magnetic charges respectively.

One derives the equations of motion by varying the above action with respect to
$a, b, c$ and $\phi$. Then we can put $c(r) = a(r)$ and the equation for $c$ turns out to be
the Hamiltonian constraint. The equations of motion and Hamiltonian constraint following from (\ref{action2}) are
\bea
&& \label{eqb} \frac{1}{3}\phi'^2 + \frac{b''}{b} =-\frac{2 G''}{b^2} + \frac{2a^2}{b^2}(G'b'^2)' \\
&& \label{eqa}(a^2b^2)'' = 2+ 8G' aa'(3a^2b'^2-1) + 8b(G'a^3a'b')'\\
&& \label{eqphi}\frac{1}{3}(a^2b^2\phi')' = \frac{1}{2b^2} \frac{dV_{eff}}{d\phi} - 2\frac{d G}{d\phi} (aa'(a^2b'^2-1))'\\
&& \label{HC}a^2b'^2-1+2aa'bb' - \frac{1}{3} a^2b^2\phi'^2 + \frac{V_{eff}}{b^2} = 4 G' a a' (3a^2b'^2-1))
\eea
where prime indicates derivation with respect to $r$.

For the attractor phenomenon to occur, it is sufficient that the following two
conditions are satisfied \cite{Goldstein:2005hq}. First, for fixed charges,
as a function of the moduli, the modified effective potential $W$, must have a critical point. Denoting
the critical values for the scalars as $\phi^i=\phi^i_0$ we have,
\bea
  \label{critical}
  \partial_i W(\phi^i_{0})=0.
\eea
Second, there should be no unstable direction around this minimum, so the matrix of second
derivatives of the potential at the critical point,
\bea
  \label{massmatrix}
  M_{ij}={1\over 2} \partial_i\partial_j W (\phi^k_{0})
\eea
should have no negative eigenvalue. Schematically we can write,
\bea
  \label{positive}
  M_{ij}>0.
\eea
We will sometimes refer to $M_{ij}$ as the mass matrix and its eigenvalues as masses (more
correctly $mass^2$ terms) for the fields, $\phi^i$. It is important to note that in deriving the conditions for the attractor phenomenon, one
does not have to use supersymmetry at all. The extremality condition puts a strong
constraint on the charges so that the asymptotic values of the moduli do not appear in the
entropy formula.

\subsubsection*{Zeroth order analysis }

Let us start by setting the asymptotic values of the scalars equal to
their critical values (independent of $r$), $\phi^i=\phi^i_0$. The
equations of motion (\ref{eqa}), (\ref{eqb}) and (\ref{eqphi}) can be easily solved.
First we solve (\ref{eqb}) and get $b(r)=r$, and then replace this
expression in (\ref{eqa}) --- we obtain:
\bea
a(r)^2=1+\frac{C_1}{r} + \frac{C_2}{r^2}
\eea
where $C_1$ and $C_2$ are integration constants. We are interested in the extremal solutions
and so the integration constants can be calculated from the `double horizon'
\footnote{The inner and outer horizons coincide and the equation has a
double root.} condition:
\be
C_1= -2r_H, \;\;\;\;\; C_2= r_H^2,
\ee
where $r_H$ is the horizon radius.
Therefore, we can write the solution as
\be
\label{a0}
a_0(r) = (1-\frac{r_H}{r}) ,
\ee
that describes the extremal RN solution. Then the dilaton equation gives rise to an important equation
\bea
\label{ma}
\frac{d V_{eff}}{d\phi} + 4 \frac{d G}{d\phi} =0
\eea
It defines the modified effective potential as $W= V_{eff} + 4 G$.

It is important to notice that, for a single charge Kaluza-Klein black hole without any higher derivative correction there is no attractor solution. This comes from the fact that attractor equation $\partial_{\phi} V_{eff}=0$ has no solution with a single charge. In this case as we have mentioned before, horizon vanishes and we have naked singularity. But in the presence of higher order correction we may find the attractor mechanism. In this case the modified attractor equation $\partial_{\phi} W=0$ may still have a solution, for an appropriate non-constant coupling  $G(\phi)$.
\bigskip

Having a solution for equation (\ref{ma}), we will find a non-zero value for the horizon radius, which means that the higher derivative  correction stretches the horizon and the naked singularity is dressed. As a result we will find non-zero value for entropy. In addition the attractor behavior is recovered. In figures (1) and (2) we compare the behavior of the scalar field for a single charged black hole, in the presence of higher order correction,  with the case without higher derivative correction. In our example we consider the coupling in higher derivative correction such as the equation  (\ref{ma}) have a solution.

\begin{figure}
 \begin{center}
  \includegraphics[scale=.5]{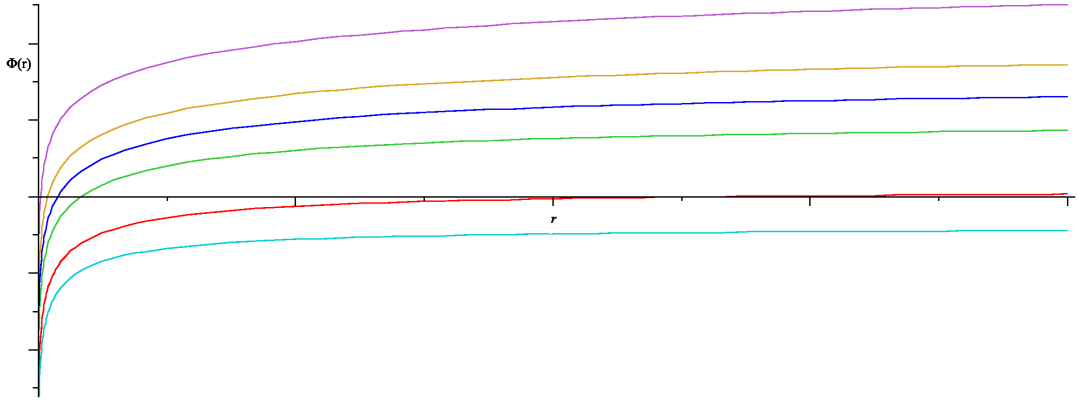}
 \end{center}
\caption{$\phi(r)$ vs. $r$, for
$Q=0$ and $P=10$. Different curves represent different asymptotic values for $\phi_{\infty}$. Horizon is located at $r_H=0$.}
\end{figure}
\begin{figure}
 \begin{center}
  \includegraphics[scale=.56]{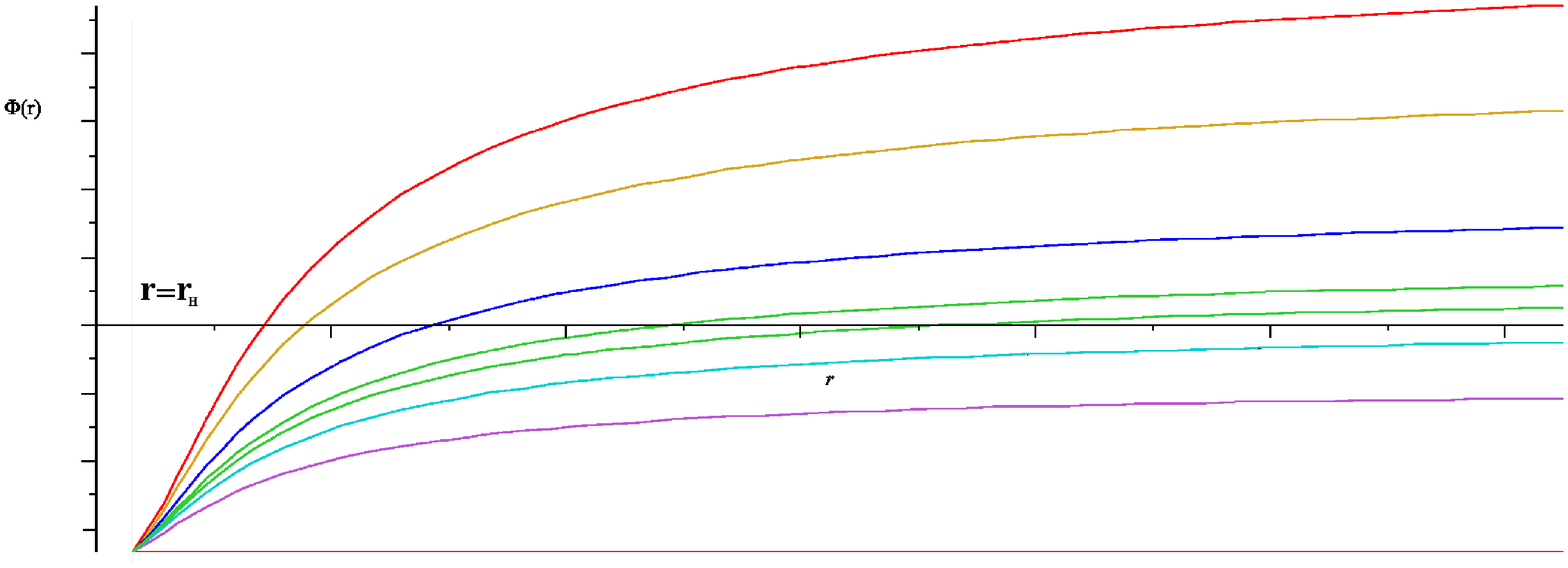}
 \end{center}
\caption{$\phi(r)$ vs. $r$, for $Q=0$ and $P=10$ with Gauss-Bonnet correction. $G(\phi)= 1/4 e^{-2\phi}$
Different curves represent different asymptotic values for $\phi_{\infty}$. The attractor point is $\phi_0=-\frac{1}{2}\ln 10$ at the horizon $r_H=\sqrt{10}$.}
\end{figure}

The Hamiltonian constraint evaluated at the boundary provides a constraint on charges. However, we are interested in solving the Hamiltonian constraint
at the horizon and in obtaining a relation between the horizon radius and the effective
potential. Using the solution for $a$ and $b$, the Hamiltonian constraint is simplified drastically at the horizon.
Thus, the horizon radius, $r_H$, is given by $V_{eff}$ at the minimum of $W$:
\bea
\label{horizonR}
r_H^2=V_{eff}(\phi_0)
\eea
It is important to notice that in this case Bekenstein Hawking entropy is not just the area of the horizon but receives a correction, which can be calculated from Weyl entropy formula. The result is proportional to the value of modified effective potential at its extremum:
\be
S_{BH} = \pi W(\phi_0)
\ee

\subsubsection*{First order analysis}
For the extremal RN black hole solution carrying the charges specified by
the parameter $Q$ and $P$ and the moduli taking the critical values $\phi_{0}$ at
infinity, a double zero horizon continues to exist for small deviations from these
attractor values for the moduli at infinity. The moduli take the critical values at the
horizon and the entropy remains independent of the values of the moduli at infinity
\cite{Goldstein:2005hq}. Now, starting with first order perturbation theory

\be
\delta \phi = \phi-\phi_0 =\epsilon \phi_1 \; ,
\ee
where  $\epsilon$  is small parameter we use to organize the perturbation theory. Using the zeroth order solution for $a$ and $b$, the first order correction to the scalar field $\phi$ satisfies the following equation

\be
\frac{1}{3} \partial_r(a_0^2b_0^2 \partial_r \phi_1) \simeq \frac{\beta^2}{2b_0^2} \phi_1
\ee
where $\beta^2$ is the second derivative of the modified effective potential $W(\phi)$ at its minimum $\phi_0$. A general solution of the second order differential equation above, is parametrized by two constants. We are interested in a smooth solution
that does not blow up at horizon $r_H$. This requirement eliminates one parameter, therefore the general smooth solution
is obtained by
\be
\phi_1 \simeq C_1 (1-\frac{r_H}{r})^{\gamma}\; ,
\ee
where $\gamma$ is the positive root of the following equation
\be
\gamma (\gamma + 1) = \frac{3 \beta^2}{2r_H^2}
\ee
and $C_1$ is an arbitrary constant, which denotes the asymptotic value of the scalar field. Asymptotically (as $r \rightarrow \infty$)  $\phi_1$ takes a constant value, $C_1$, however it vanishes at
the horizon and the value of the scalar is fixed at $\phi_0$ regardless of $C$, its value at infinity.
It is clear that if the $\beta^2$ is positive, then the solution is regular at the horizon and so the existence of a regular horizon is related to the existence of the attractor mechanism.

\bigskip
We should also notice that, for a general coupling,  $\partial_{\phi} G(\phi_0)$ does not vanish when $W$ and $G$ does not have the same extremum. Therefore from equations (\ref{eqb}), (\ref{eqa}) and (\ref{HC}) we find that $a(r)$ and
$b(r)$ receive a correction of order $\epsilon$. In this case, in order to find the first order correction we need to solve all four equations (\ref{eqb})-(\ref{HC}) together.

\subsubsection*{Higher order analysis}

Going to higher orders in perturbation theory is in principle straightforward.  We solve the system of equations (\ref{eqb}),(\ref{eqa}),(\ref{eqphi}) and (\ref{HC}) order by order in the $\epsilon$
expansion.

To the first order, we find that one variable, say $C_1$ , can not be fixed by the
equations.  Thus we will find $a_1(r), b_1(r), \phi_1(r)$ as a function of $C_1$.
One can check that at any order $n > 2$, we can substitute the resulting values
of $(a_m , b_m )$, for all $m <n$  from the previous orders. Then equations (\ref{eqb})-(\ref{HC})
of the current order together, consistently give,
\be
a_n = a_n(C_1 )\;, \;\;\;\;\; b_n = b_n (C_1 ) \;, \;\;\;\;\;     \phi_n = \phi_n(C_1)
\ee
as polynomials of order $n$ in terms of $C_1$.
     It is worth noting that $C_1$ remains a free parameter to all orders in the $\epsilon$-
expansion.  Owing to the result above, we observe that $(a_{\infty},  b_{\infty}, \phi_{\infty} )$ are
varying and will take different values, for different choices for $C_1$.
The arbitrary value of $\phi$ at infinity is $\phi=\phi_{\infty}$  , while its value at the horizon is fixed
to be $\phi_0$ .

\subsection{Kaluza-Klein Black Holes with general $R^2$ corrections}

 To study the effect of higher derivative corrections, here we apply the entropy function formalism to static black holes solution. It was shown by Sen that the attractor mechanism is related to the
extremality rather than to the supersymmetry property of a given solution.
Therefore, the condition for the existence of the attractor mechanism
is the existence of an $AdS_2$ as part of the near horizon geometry of an extremal black hole. The entropy function is defined as
\bea
  {\mathcal E}({u},{v},{e},
  {p})=
  2\pi\bigg(eq-f({u},{v},
  {e},{p})\bigg)=
  2\pi\bigg(eq-\int d\theta d\varphi \sqrt{-G}{\cal L}\bigg),
\eea
where $Q=\partial f/\partial e$ is the electric charge, $u$ is the value of the
moduli at the horizon, and ${e}$ is the
electric field and $v_1$, $v_2$ are the sizes of $AdS_2$ and $S^2$, respectively. Thus,
$\mathcal{E}/2\pi$ is the Legendre transform of the function $f$
with respect to the variables $e$. The reason why it is not a Legendre
transform with respect to magnetic charge is due to topological character of the
magnetic charge. The Bianchi identities do not change when the action is supplemented
with $\alpha'$-corrections, but the equations of motion receive corrections. Then it follows
as a consequence of the equations of motion that for a black hole
carrying electric charge $Q$ and magnetic charge $P$, the constants $v_1, v_2, u$ and $e$ are given by:

\bea
  \frac{\partial {\mathcal E}}{\partial u}=0\,, \qquad
  \frac{\partial {\mathcal E}}{\partial v_i}=0\,,
  \qquad \frac{\partial {\mathcal E}}{\partial e}=0\,.
  \label{attractor}
\eea

Furthermore, the entropy associated with the black hole is given by $ S_{BH}= {\mathcal E}({u},{v},{e},{p})$ at the
extremum (\ref{attractor}). If ${\mathcal E}$ has no flat directions, then the
extremization of ${\mathcal E}$ determines ${u}$, ${v_i}$,
${e}$ in terms of ${Q}$ and ${P}$. Therefore,
$S={\mathcal E}$ is independent of the asymptotic values of the scalar fields. These results lead to
a generalized attractor phenomenon for both supersymmetric and non-supersymmetic extremal
black hole solutions.

We can apply this method to our action (\ref{action1}).
The general metric of near horizon geometry $AdS_2\times S^2$ can be written as
\bea
  ds^2=v_1(-\rho^2d\tau^2+\frac{1}{\rho^2}d\rho^2)+
  v_2d\Omega_2^2\, .
  \label{adshor}
\eea
The field strength ansatz in our case is given by
\bea
  F=ed\tau\wedge d\rho + P\sin\theta d\theta\wedge d\varphi.
\label{gh}
\eea
For the metric  (\ref{adshor}) Ricci scalar is $R=2/v_2-2/v_1$. By dimensional argument it is clear that for the metric (\ref{adshor}) a general $R^2$ correction term like (\ref{R2c}) is given by
\be
{\mathcal L}_2 = \frac{c_1}{v_1^2} + \frac{c_2}{v_1 v_2} + \frac{c_3}{v_2^2}
\ee
where comparing this with (\ref{R2c}) we will find that
\be
c_1=c_3=4\alpha + 2\beta + 4\gamma \;,;\;\;\;\;\; c_2=-8\alpha\;.
\ee

The entropy function ${\mathcal E}(u,v_1,v_2,Q,P)$ and $f(u,v_1,v_2,e,P)$ are given by\footnote{We set the Newton's constant $G_4=1$}
\bea
\label{Fmisto}
  && {\mathcal E} (u,v_1,v_2,Q,P)=2\pi [Q e-f(u,v_1, v_2, e)]\, ,\\
  \nonumber
  && f(u,v_1,v_2,e,P)=\frac{1}{2}\left( v_1-v_2 -(\frac{P^2}{v_2^2}-\frac{e^2}{v_1^2})v_1v_2  e^{2u} + G(u)(\frac{c_1v_2}{2v_1} + \frac{c_2}{2} +\frac{c_3v_1}{2v_2} )\right)\, .
\eea
Then the attractor equations are obtained as :
\bea
  \label{at1}
  \frac{\partial {\mathcal E}}{\partial v_1} & = & 0\,\,\,\Rightarrow
  \,\,\,-1+ \frac{1}{v_2}V_{eff}(u)+\left(\frac{c_1v_2}{2v_1^2} - \frac{c_3}{2v_2}\right) G(u)=0   \, ,\\
  \label{at2}
  \frac{\partial {\mathcal E}}{\partial v_2} & = & 0\,\,\,\Rightarrow \,\,\,1-\frac{v_1}{v_2^2} V_{eff}(u)-\left(\frac{c_1}{2v_1} - \frac{c_3v_1}{2v_2^2}\right) G(u)=0\, ,\\
  \label{at3}
  \frac{\partial {\mathcal E}}{\partial u} & = & 0\,\,\,\Rightarrow \,\,\,
   \,\frac{v_1}{v_2}  \frac{d V_{eff}(u)}{d u} -     \left(\frac{c_1v_2}{2v_1} + \frac{c_2}{2} +\frac{c_3v_1}{2v_2} \right)      \frac{d G(u)}{d u} =0 ,\\
  \label{at4}
  \frac{\partial {\mathcal E}}{\partial e} & = & 0\,\,\,\Rightarrow \,\,\,
Q-\frac{e v_2}{v_1}\; e^{2 u}=0\, ,
\eea
Solving the first two equations gives us $v_1=v_2=V_{eff}$. Therefore the third equation is:
\bea
\label{eqA}
 \frac{d V_{eff}(u)}{d u} - \frac{c}{2}\frac{d G(u)}{d u} =0\; ,
\eea
where $c=c_1+c_2+c_3$.  This is an important equation. It is equivalent to finding the critical points of the modified effective potential $W=V_{eff}- cG$
at horizon. For the Gauss-Bonnet correction where $c_1=c_3=0, c_2 =-8$, this is the same potential we have found in our previouss analysis.
As we have shown for Gauss-Bonnet case, the equation (\ref{eqA}) is one of the conditions for the existence of attractor mechanism.
If this equation has solutions, then the moduli values at the horizon are fixed in term of
the charges.

\bigskip

As we have discussed before,  while for a single charged Kaluza-Klein horizon vanishes and we do not have the attractor mechanism, in the presence of higher derivative correction moduli may be fixed at horizon due to extra term in modified effective potential: Higher derivative correction stretches the horizon and the naked singularity will be dressed. As a result we will find non-zero value for entropy. Although for single charged black hole in this theory, the necessary condition for having a solution for equation (\ref{eqA}) is having a non-constant coupling $G(\phi)$. 

\bigskip

Finally it is important to notice that the existence of a near-horizon
geometry when the moduli are not constants does not imply the existence of the
whole solution in the bulk (from the horizon to the boundary) this is the disadvantage
of the entropy function formalism. However, in the next subsection we will investigate
the equations of motion in the bulk and describe the horizon as an IR critical
point of the effective potential.

\bigskip

By replacing $v_1=v_2=V_{eff}(\phi_0)$ where $\phi_0$ is the solution of equation $\partial W=0$ into the (\ref{Fmisto}), we
obtain the value of the entropy function at the extremum:

\bea
\label{coren}
S_{BH}=\pi W(\phi_0)\;.
\eea
As an example of $R^2$ corrections to Kaluza-Klein black hole, let us consider $R^2$ terms in the dimensional reduction of type $IIA$ string theory. One way to get such correction is starting with five-dimensional original solution, taking the product  with a flat $T^6$  we obtain a solution to M-theory whose type $IIA$ reduction has $D0$ and $D6$ charges. The first correction to the effective action of type $IIA$ theory involves terms with eight derivatives including $R^4$ terms, and derivative of dilaton and $NS-NS$ and $RR$ fields. Here we restrict ourselves to the $R^4$ terms. This includes
\bea
\label{corretion1}
&& S_{tree} = \frac{1}{16\pi G_{10}} \int dx^{10}  \sqrt{-g} e^{-2\phi}\left( R+\frac{\alpha'^3 \zeta(3)}{3.2^6} (t_8t_8 +\frac{1}{8}\epsilon_{10}\epsilon_{10} )(R)^4 \right) \\
\label{corretion2}
&& S_{1-loop} =  \frac{1}{16\pi G_{10}} \frac{\alpha'^3 \pi^2}{9.2^6}g_s^2\int dx^{10}  \sqrt{-g} e^{-2\phi}(t_8t_8 -\frac{1}{8}\epsilon_{10}\epsilon_{10} )(R)^4
\eea
In the equations above we have used the following standard notation to denote index contractions:
\be
\epsilon_{10}\epsilon_{10} R^4\equiv \epsilon^{\alpha\beta\mu_1\nu_1\ldots\mu_4\nu_4}
\epsilon_{\alpha\beta\rho_1\sigma_1\ldots\rho_4\sigma_4}\,R^{\rho_1\sigma_1}_{\mu_1\nu_1}\ldots
R^{\rho_4\sigma_4}_{\mu_4\nu_4}
\ee
\be
t_8 t_8 R^4 \equiv  t_8^{\mu_1\nu_1\ldots\mu_4\nu_4}
t_{8\,\rho_1\sigma_1\ldots\rho_4\sigma_4}\,R^{\rho_1\sigma_1}_{\mu_1\nu_1}\ldots
R^{\rho_4\sigma_4}_{\mu_4\nu_4}
\ee
where the $t_8$ tensor is defined as ($F^{(i)}_{\mu\nu}$, $i=1,\ldots,4$ are any anti-symmetric tensors) 
\bea
&&t_8^{\mu_1\nu_1\ldots\mu_4\nu_4} F^{(1)}_{\mu_1\nu_1}\ldots F^{(4)}_{\mu_4\nu_4}\nonumber\\
&&= 8 (F^{(1)}_{\mu\nu}F^{(2)\,\nu\rho} F^{(3)}_{\rho\lambda} F^{(4)\,\lambda\mu}+
F^{(1)}_{\mu\nu}F^{(3)\,\nu\rho} F^{(2)}_{\rho\lambda} F^{(4)\,\lambda\mu}+
F^{(1)}_{\mu\nu}F^{(3)\,\nu\rho} F^{(4)}_{\rho\lambda} F^{(2)\,\lambda\mu})\nonumber\\
&& -2 (F^{(1)\,\mu\nu}F^{(2)}_{\mu\nu}F^{(3)\,\rho\lambda}F^{(4)}_{\rho\lambda}+
F^{(1)\,\mu\nu}F^{(3)}_{\mu\nu}F^{(2)\,\rho\lambda}F^{(4)}_{\rho\lambda}+
F^{(1)\,\mu\nu}F^{(4)}_{\mu\nu}F^{(2)\,\rho\lambda}F^{(3)}_{\rho\lambda})
\eea

Now by reducing the above action on $K3\times T^2$ we get both $R^2$ and $R^4$ correction terms in the four-dimensional effective action \cite{Harvey:1996ir}, where $R^2$ term is given by

\be
{\mathcal L}_2 = 
{C_2\over 6}\,{g_s^2\,\alpha'^3\over V_0}\,{T\over T_0}\,
R_{\mu\nu\rho\sigma}R^{\mu\nu\rho\sigma} \, .
\ee
The volume of $K3$ and $T^2$ are $(2\pi)^4 V$ and $(2\pi)^2 T$ respectively and $V_0$ and $T_0$ are the value of $V$ and $T$ at infinity. $C_2$ is the second chern class of $K3$ which is $24$. The ten-dimensional type $IIA$ metric is given by 
\be 
ds_{10}^2 = ds_4^2 + e^{2\phi/3} \sum_{i=1}^6 dy_i dy^i
\ee   
therefore
\be 
\frac{T}{T_0} = e^{2\phi/3}
\ee 
For this correction term, $c_2=0, c_1=c_3=\frac{16 g_s^2\alpha'^3}{V_0} $ therefore the effective potential is given by
\be 
W= Q^2 e^{-2\phi} + P^2e^{2\phi} - \frac{16 g_s^2\alpha'^3}{V_0} e^{2\phi/3}
\ee 
Minimizing this potential gives us
\be
e^{2\phi_0} = |\frac{Q}{P}| \left(1+ \frac{8 g_S^2 \alpha'^3}{3V_0P^{4/3}Q^{2/3}} + {\mathcal O}(\alpha'^6)\right) 
\ee
therefore the first correction to the entropy which is given by (\ref{coren}) is
\be 
S_{BH} = 2\pi |PQ| -  \frac{16 g_S^2 \alpha'^3}{V_0} |\frac{Q}{P}|^{1/3} \;.
\ee 
From (\ref{at1}) and (\ref{at2}) we find that the radii of $AdS_2$ and $S^2$ do not receive any correction in order ${\mathcal O}(\alpha'^3)$. 
\section{Kaluza-Klein Black Holes with general higher derivative corrections}

Our results in the previous section, $R^2$ corrections to the attractor mechanism for Kaluza-Klein, can be generalized to higher order corrections. To start let us consider a general $R^4$ correction to the action (\ref{action0}). We are interested in studying static and extremal black hole solution with $SO(2,1)\times SO(3)$ invariant near horizon geometry. Near horizon metric (\ref{adshor}) and gauge field (\ref{gh}) are the most general possible forms, consistent with this symmetry. Using this near horizon geometry it is easy to show that, at the horizon the most general $R^4$ correction has the following structure
\be
{\mathcal L}_4 = \frac{c_1}{v_1^4} + \frac{c_2}{v_1^3v_2}+\frac{c_3}{v_1^2v_2^2}+\frac{c_4}{v_1v_2^3}+\frac{c_5}{v_2^4}
\ee
where constants $c_1$-$c_5$ are fixed in term of numerical factors in the $R^4$ structure. Using the definition  (\ref{Fmisto}), it is easy to calculate  the entropy function with this additional term. It is given by
\be
{\mathcal E} = \pi \left[v_2-v_1 + \frac{v_1}{v_2} V_{eff}(u) - \frac{\gamma(u)}{2} v_1v_2 (\frac{c_1}{v_1^4} +\frac{c_2}{v_1^3v_2} + \frac{c_3}{v_1^2v_2^2} + \frac{c_4}{v_1 v_2^3} + \frac{c_5}{v_2^4}   ) \right]
\ee
The value of $v_1$ and $v_2$ at the horizon are determined by extremizing $\mathcal E$ with respect to $v_1$ and $v_2$. This gives
\bea
  \label{att41}
&& -1+ \frac{1}{v_2}V_{eff}(u) + \gamma(u)  \left(\frac{3c_1 v_2 }{2v_1^4}+\frac{c_2  }{v_1^3}+ \frac{c_3}{2v_1^2v_2} -\frac{c_5}{2v_2^3}\right)=0 \; ,\\
  \label{att42}
&&\;\;\;1-\frac{v_1}{v_2^2} V_{eff}(u) + \gamma(u)  \left(-\frac{c_1  }{2v_1^3}+ \frac{c_3}{2v_1v_2^2}+\frac{c_4}{v_2^3} +\frac{3c_5v_1}{2v_2^4}\right)=0 \; .
\eea

Now we can solve these two equations to find $v_1$ and $v_2$.  For small coupling $\gamma$ we find

\bea
\label{v1}
&&v_1= V_{eff} \left(1+ \frac{5c_1  +4c_2+3c_3+ 2c_4+c_5}{2V_{eff}^{3}} \; \gamma+ \mathcal{O}(\gamma^2)  \right)\; , \\
\label{v2}
 &&v_2= V_{eff} \left(1+ \frac{3c_1  +2c_2+c_3-c_5}{2V_{eff}^{3}} \; \gamma+ \mathcal{O}(\gamma^2)  \right)\; .
\eea

Scalar field equation in the near horizon geometry correspond to extremizing the entropy function with respect to $u$, which gives us
\be
  \label{att3}
 \frac{v_1}{v_2}\frac{d V_{eff}}{d u} -\frac{1}{2}\frac{d \gamma}{du} v_1v_2 \left(\frac{c_1  }{v_1^4}+  \frac{c_2  }{v_1^3v_2}+\frac{c_3}{v_1^2v_2^2}+\frac{c_4}{v_1v_2^3} +\frac{c_5}{v_2^4}\right)=0 \; .
\ee

Using equations (\ref{att41}) and (\ref{att42}) we can simplify the scalar equation into the following form

\be
\label{eff1}
 \frac{v_1}{v_2}\frac{d V_{eff}}{d u} +\frac{v_2-v_1}{2\gamma}\frac{d \gamma}{du} =0 \; ,
\ee
and finally using the solutions (\ref{v1}) and (\ref{v2}), for small coupling $\gamma$ we  find :

\be
 \frac{d}{du} \left( V_{eff} -\frac{c\;\gamma}{2V_{eff}^2}\right) \simeq 0\; ,
\ee
where $c=c_1+... +c_5$. This equation is analogous to the attractor equation (\ref{eqA}), which we have found for $R^2$ correction in the previous section. Therefore it suggests defining the modified effective potential at the horizon as:
\bea
\label{modeff}
 W= V_{eff} -\frac{c\;\gamma}{2V_{eff}^2} \;.
\eea
In addition if we consider dilaton equation at the near horizon geometry (\ref{adshor}), we will find that the similar term appears on the right hand side of the dilaton equation. Therefore our definition for modified effective potential makes sense. By the same analysis we did in the section 3, we can find that the condition of having attractor solution is reduced to having a minimum for the modified effective potential $W$
\be
\label{atc}
\partial W(\phi_0) =0 \;, \;\;\;\; \partial^2 W(\phi_0) > 0 \; .
\ee
The black hole entropy is given by the value of entropy function at its extremum. It is easy to show that this is proportional to the value of modified effective potential at its minimum

\bea
\label{ModEnt}
S_{BH}= \pi W(\phi_{0i}) \;.
\eea

As an example for $R^4$ correction, let us consider $R^4$ terms in the dimensional reduction of type $IIA$ string theory. As we said, the first correction to the effective action of type $IIA$ theory involves terms with eight derivatives including $R^4$ terms (\ref{corretion1}, \ref{corretion2} ). Now if we reduce this action on $T^6$ we get both $R^4$ corrections terms in the four-dimensional effective action. Consider the near horizon geometry (\ref{adshor}), the effect of $R^4$ corrections (\ref{corretion1}, \ref{corretion2} ) to four-dimensional effective action is given by

\be 
{\mathcal L}_4 = \frac{\gamma(u)}{108} \left( \frac{35}{v_1^4}-\frac{5}{v_1^3v_2} + \frac{21}{v_1^2v_2^2} -\frac{5}{v_1v_2^3} + \frac{35}{v_2^4}\right)\;,
\ee
where coupling $\gamma(u)$ is given by
\be
\gamma(u) = \gamma_0 e^{2 u} = 4 (\zeta(3) + \frac{\pi^2g_s^2}{3}) \alpha'^3 e^{2u} \, .
\ee 

By minimizing the modified effective potential (\ref{modeff}) we can find the attractor value $\phi_0$
\be
e^{2\phi_0} = |\frac{Q}{P}| + \frac{3 \gamma_0}{64 |Q P^5|} + {\mathcal O}(\gamma_0^2) \, ,
\ee
and the correction to the entropy from equation (\ref{ModEnt}) is given by

\be 
S= 2\pi|PQ| -\frac{3 \pi \gamma_0}{32|QP^3|} + {\mathcal O}(\gamma_0^2) \,.
\ee 
Finally the corrections to the $AdS_2$ and $S^2$ radii are given by

\be
v_1= 2 |P Q| + \frac{9\gamma_0}{32 P^2 Q^2} + {\mathcal O}(\gamma_0^2)\; , \;\;\;\;\; v_2= 2 |P Q| + \frac{3\gamma_0}{32 P^2 Q^2} + {\mathcal O}(\gamma_0^2) \; .
\ee

 It is straightforward to repeat the above calculus for a general $R^n$ correction. By dimensional analysis one can show that, at the near horizon the most general $R^n$ correction has the following structure
\bea
\label{cl}
{\mathcal L}_n = \frac{c_1}{v_1^n} + \frac{c_2}{v_1^{n-1}v_2}+...\frac{c_{n+1}}{v_2^n}
\eea
Then the entropy function is given by
\be
{\mathcal E} = \pi \left[v_2-v_1 + \frac{v_1}{v_2} V_{eff}(u) - \frac{\gamma(u)}{2} v_1v_2 \sum_{i=0}^{n} \frac{c_{i+1}}{v_1^{n-i}v_2^{i}} \right]
\ee
By extremizing above entropy function, we will find the values of $v_1$ and $v_2$ as follows

\bea
\label{vn1}
&& \hspace{-1cm}v_1= V_{eff} \left[1+ \frac{1}{2V_{eff}^{n-1}} \bigg((2n-3)c_1 + (2n-4)c_2 + ... (n-3)c_{n+1}\bigg) \; \gamma+ \mathcal{O}(\gamma^2)  \right] , \\
\label{vn2}
&&  \hspace{-1cm} v_2= V_{eff} \left[1+ \frac{1}{2V_{eff}^{n-1}} \bigg( (n-1)c_1 + (n-2)c_2 + ... +c_{n-1}-c_{n+1}\bigg) \; \gamma+ \mathcal{O}(\gamma^2)  \right] ,
\eea
and the resulting  scalar field equation is given by

\bea
\label{mae}
 \frac{d}{du} \left( V_{eff} -\frac{c\;\gamma}{2V_{eff}^{n-2}}\right) \simeq 0\; ,
\eea
where we defined $c=c_1+...+c_{n+1}$ which is the correction Lagrangian (\ref{cl}) evaluated for the geometry (\ref{adshor}) with $v_1=v_2=1$. Finally we can define the modified effective potential $W$ for a general $R^n$ correction as follows
\be
W= V_{eff} -\frac{c\;\gamma}{2V_{eff}^{n-2}}.
\ee
By the same argument we can find that the attractor conditions are (\ref{atc}). The entropy of the black hole is given by the entropy function at its extremum. If we denote the solution of equation (\ref{mae}) by $\phi_0$, the black hole entropy which is the value of entropy function at its extremum is given by
\be
S_{BH}= \pi \left[ V_{eff}(\phi_0)-\frac{c\;\gamma(\phi_0)}{2V_{eff}^{n-2}(\phi_0)} \right]\;,
\ee
which is equal to the value of modified effective potential at the horizon. Our discussion can be easily generalized for any extremal black hole solution in Einstein-Hilbert gravity coupled with arbitrary number of $U(1)$ gauge fields and arbitrary number of moduli fields with a general coupling between gauge field and scalar. It can be described by the following action
\bea
\label{ga}
I= \frac{1}{16 \pi} \int d^4x \sqrt{-G}\bigg(&& R-2\partial^{\mu}\Phi_i \partial_{\mu}\Phi^i- f_{ab}(\phi_i) F^{a\;\mu\nu}F^{b}_{\;\mu\nu} \cr\cr && -\frac{1}{2} {\tilde f}_{ab}(\phi_i)\epsilon^{\mu\nu\alpha\beta}F^{a\;\mu\nu}F^{b\;\mu\nu} + \gamma(\phi_i) {\mathcal L}_n\bigg) \; ,
\eea
where $f_{ab}(\phi_i)$ and $\tilde{f}_{ab}(\phi_i)$ determine the general moduli dependent
gauge couplings. Then one can show that the effective potential is given as follows \cite{Goldstein:2005hq}:

\be
V_{eff}(\phi_i) = f^{ab}(Q_a-\tilde{f}_{ac}P^c)(Q_b-\tilde{f}_{bd}P^d) + f_{ab}P^aP^b
\ee
where $Q^a$ and $P^a$ are denoted electric and magnetic charges. In the absence of higher derivative corrections, it has been shown that there are two conditions which are sufficient for the existence of
an attractor solution\cite{Goldstein:2005hq}. First, the charges should be such that the resulting effective
potential, $V_{eff}(\phi_i)$, has a critical point. We denote the critical
values for the scalars as $\phi_i = \phi_{i0}$. So that,
\be
\partial_i V_{eff}(\phi_{i0})=0
\ee
Second, the matrix of second derivatives of the potential at the critical point,
\be
M_{ij} = \frac{1}{2} \partial_{ij} V_{eff}(\phi_{0i})
\ee
should have positive eigenvalues. Once these two conditions hold, we show below that the attractor phenomenon results. Since our results in the previous section are independent of explicit form of effective potential $V_{eff}$, we can simply generalize our results to the extremal black hole solution of the action (\ref{ga}). Therefore the modified effective action is given by
\be
W(\phi_i)= V_{eff}(\phi_i) -\frac{c\;\gamma(\phi_i)}{2V^{n-2}_{eff}(\phi_i)},
\ee
The conditions of having attractor solution are changed to having a critical point for modified effective potential $W(\phi_i)$ and positivity of the matrix of second derivatives of the modified potential at the critical point,
\be
\partial_{i} \partial_j  W(\phi_{0i}) > 0 \;.
\ee
And finally the entropy which is given by the value of the entropy function at its extremum is

\be
S_{BH}= \pi W(\phi_{0i}) \;,
\ee
which is equal to the value of the effective potential $W$ at its critical point $\phi_{i0}$.

\section{Discussion}

In this paper, we have investigated the attractor solution for Kaluza-Klein black hole in the presence of higher derivative terms. We started with the simplest higher derivative correction, i.e. Gauss-Bonnet term. Although in four dimensions this is a total derivative but since we considered it coupled to the scalar field we can not integrate it out and it affects the solution. By investigating solutions of the equations of motion, we observed the attractor
behavior explicitly. We looked for all possible solutions which admit the criteria of being regular at the horizon and free in the asymptotic region. The near horizon analysis shows the criteria for attractor behavior. We introduced the modified effective potential $W$, which plays the role of the scalar effective potential introduced in \cite{Goldstein:2005hq}, in the presence of higher derivative correction. The conditions for having the attractor solution can be expressed in terms of two conditions on this function. Moreover, the entropy is proportional to this function at its extremum. We extended our studying to the more general higher derivative corrections $R^n$. For small scalar coupling $\gamma$, we have found the modified the effective potential as well.

\bigskip

There is a simple embedding of the Kaluza-Klein black hole into string theory. Taking the product of extremal Kaluza-Klein black hole solution with a flat $T^6$ (or a general Calabi-Yau three-fold) , we obtain a solution of $M$ theory, whose $IIA$ reduction has $D0$ and $D6$ charge. In \cite{Emparan:2006it} authors have found the microscopic origin for Kaluza-Klein black hole entropy. The idea is based on the fact this some neutral
black holes can be lifted to M-theory in such a way that the reduction to $IIA$
string theory has both $D0$ and $D6$ charge. Then one can count the number
of $D0-D6$ bound states. To count the bound states, suppose $N0 = 4 k^3 N, N6 = 4l^3 N$. If we consider the $T^6$ as a product of three $T^2$’s and T-dualize along one cycle of each $T^2$, we get a configuration
of four stacks of $D3$-branes wrapping the diagonal cycles of the $T^2$’s. There
are $N$ branes in each stack. If the $D3$-branes were wrapping the fundamental
cycles instead, this configuration would be equivalent to a four charge black
hole whose microscopic entropy per intersection is known to be $2\pi N^2$. There are now $(2kl)^3$ intersection points in total,  therefore the entropy is given by $ 2\pi (2kl)^3  N^2$. Using the relation between electric and magnetic charges of KK black and number or $D0$ and $D6$-branes we get the correct expression for the black hole entropy. This was shown to work for both static and rotating Kaluza-Klein black holes as long as they had sufficient $D0$ and $D6$ charges after dimensional reduction. Having a successful statistical description for  Kaluza-Kleins from string theory, it would be quite interesting if one can find the leading order correction to statistical entropy of the Kaluza-Klein and compare it with our results from gravity side.

\section*{Acknowledgments}
We would like to thank Dumitru Astefanesei, Kevin Goldstein, Sangmin Lee and Soo-Jong Rey for discussions.
This work is supported by the Korea Research Foundation Leading Scientist Grant
(R02-2004-000-10150-0) and Star Faculty Grant (KRF-2005-084-C00003).

\end{document}